\documentclass[preprint,noshowpacs,showkeys,preprintnumbers,amsmath,amssymb]{revtex4}

\usepackage{graphicx}
\usepackage{dcolumn}

\begin{document}
\preprint{APS/123-QED}

\title{A survey of fractured SrTiO$_{3}$ surfaces: from the micro-meter to nano-meter scale}

\author{TeYu Chien$^{1}$}
\email{tchien@anl.gov}
\author{Nathan P. Guisinger$^{2}$}
\author{John W. Freeland$^{1}$}%
\affiliation{
$^{1}$Advanced Photon Source, Argonne National Laboratory, Argonne, IL 60439\\
$^{2}$Center for Nanoscale Materials, Argonne National Laboratory, Argonne, IL 60439
}%

\date{\today}

\begin{abstract}
Cross-sectional scanning tunneling microscopy was utilized to study fractured perovskie oxide surfaces. It was found for the non-cleavable perovskite oxide, SrTiO$_{3}$, that atomically flat terraces could be routinely created with a controlled fracturing procedure. Optical and scanning electron microscopy as well as a profilometer were used to obtain the information from sub-millimeter to sub-micrometer scales of the fractured surface topography.
\end{abstract}

\maketitle
Perovskite oxide materials have drawn plenty attention due to their abundant phases which can be tuned by altering chemical doping and/or different external stimuli. For example, pure SrTiO$_{3}$(STO) is a paraelectric insulator but when doped with small amount of Nb or oxygen vacancies, it converts to a high-mobility metal \cite{yamada 1973}. It was also demonstrated that the room temperature (RT) ferroelectric phase could be achieved in a strained thin film \cite{haeni 2004}. Recently, interfaces of epitaxial perovskite oxides exhibited many new phases not observed in the bulk constituents on either side of the interfaces. Examples are: spin rearrangement, orbital reconstruction and covalent bonding at interface between La$_{2/3}$Ca$_{1/3}$MnO$_{3}$ and YBa$_{2}$Cu$_{3}$O$_{7}$ \cite{chakhalian 2006, chakhalian 2007}; superconducting interface between cuprate insulator and metal \cite{gozar 2008}; and highly mobile 2D electron gas at the interface of insulating LaAlO$_{3}$ and SrTiO$_{3}$ \cite{ohtomo 2004}. Though many reports pointed out the perovskite oxide interfaces could exhibit new phases, a mature direct local probe of the electronic properties right at the interface is still lacking. It has been demonstrated that the cross-sectional scanning tunneling microscopy (XSTM) can be used to map the band bending across a semiconductor interface for more than a decade \cite{gwo 1993}. Recently, Basletic \textit{et al.} demonstrated that the metallic states localized at an oxide interface could be probed with the conducting atomic force microscope on sample cross-sections that were mechanically polished \cite{basletic 2008}. Here we demonstrate that fracturing the oxide samples routinely results in atomically flat surfaces. The morphology survey shown here infers that the capabilities of XSTM with high spatial resolution imaging, while concurrently probing the local density of states, at oxide interfaces to be very promising.

It is well known that the perovskite oxide materials do not have a well-defined cleavage plane. Thus, cleaving is not an accurate way to describe our cross-sectional studies, but rather that the samples are fractured. Traditionally, for STO, to prepare fresh surfaces, chemical etching or annealing in different environments were commonly used (see ref. \cite{bonnell 2008} and references therein). Recently, we have achieved preliminary STM studies of pristine Nb-doped STO (Nb:STO) surfaces by a novel fracturing procedure \cite{guisinger 2009, chien 2009}. Here, we summarized the possible surface morphologies observed in the samples fractured with this novel procedure.

All of the samples tested here are commercial Nb:STO, purchased from Crystec. The sample dimensions used in our experiments can be mainly sorted into two categories with size of either $\sim$5.0$\times$2.0$\times$0.5 mm$^{3}$ or $\sim$10$\times$2.5$\times$1.0 mm$^{3}$. The samples were scribed from back or from side, in order to vary the fracturing directions from [001] or from [100], respectively. The scribed samples were mounted vertically in the sample holder with the scribed line aligned to the top of the clamp to maximize the sample stability during fracturing. After the fresh surfaces were created, the STM tip was introduced to scan the fresh fractured surfaces. For a schematic, please refer to ref. \cite{guisinger 2009}.

\begin{figure}
\includegraphics[width=1\textwidth]{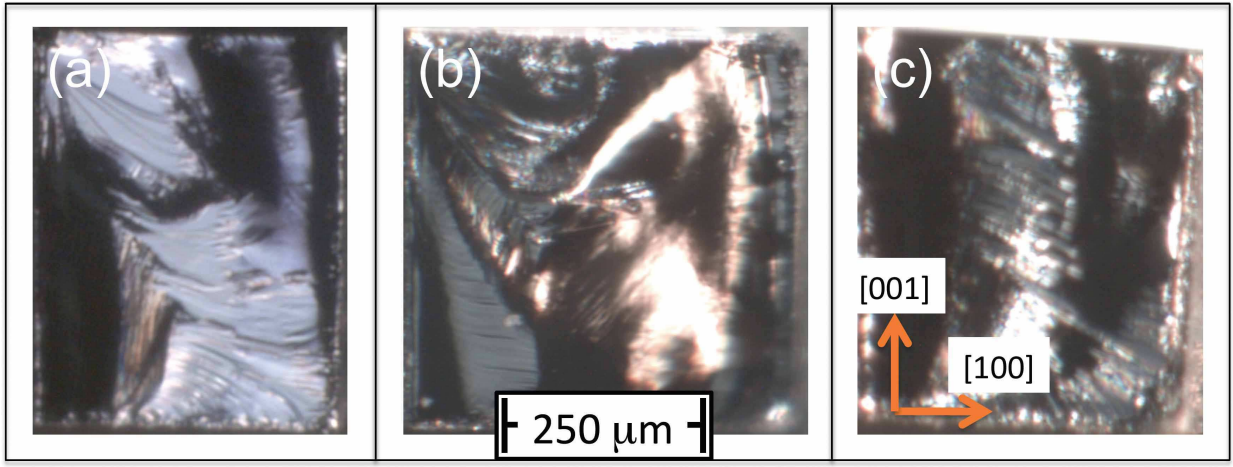}
\caption{\label{fig:OM} Optical microscope images of three difference fractured surfaces of Nb:SrTiO$_{3}$. The sample orientations for all three samples are indicated in (c).}
\end{figure}

Figure \ref{fig:OM} shows a collection of optical microscope (OM) images for three different fractured Nb:STO surfaces.
These samples were 0.5 mm thick and were fractured along [100] direction. First observation is that the fractured surfaces have similar fracture pattern: sub-millimeter wavy textures along the [001] (perpendicular to the fracturing direction). Second, it is obvious that the fractured surfaces are not globally flat. This is the signature of the conchoidal fracture, which has been seen in studies of fractured perovskite materials and many other amorphous or glass materials \cite{ahmad 2005}. The conchoidal fracture is commonly seen in brittle materials, where the materials do not possess a plastic deformation behavior. This also reflects the fact that the perovskite crystal structure does not possess cleavage plane. 

\begin{figure}
\includegraphics[width=1\textwidth]{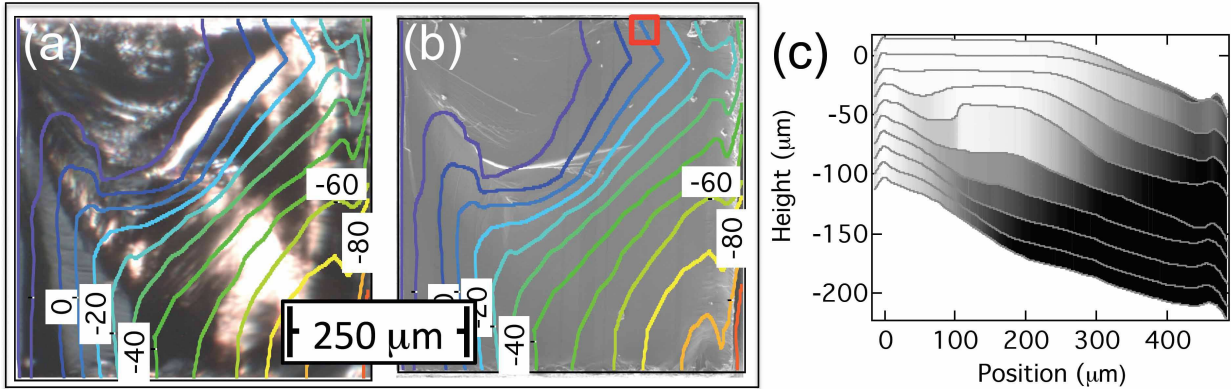}
\caption{\label{fig:contour} (a) OM picture; (b) SEM image of one fractured surface of Nb:SrTiO$_{3}$. (c) the profilometer data of the same surface. Constant height contour plot according to the profilometer data was added in (a) and (b). The numbers shown in contour plot are in the unit of $\mu$m.}
\end{figure}

Figure \ref{fig:contour} (a) and (b) show the OM and scanning electron microscope (SEM) images at the same magnification of the sample shown in fig. \ref{fig:OM} (b), respectively.
In the SEM image, fig. \ref{fig:contour} (b), one bright curved feature was observed. The bright feature observed in SEM image originates from the edge effect, which comes from an abrupt height change. To quantify the vertical scale, profilometer measurements were taken (see fig. \ref{fig:contour} (c)). Each curve was measured along [100] direction over the whole fractured region with 50 $\mu$m a part along [001] direction. A constant height contour plot from this data of fig. \ref{fig:contour} (c) was overlaid in fig. \ref{fig:contour} (a) and (b). It revealed that the bright feature in SEM is due to the height change about 10 $\mu$m in a lateral scale of $\sim$10 $\mu$m, which is about 45 degree incline. The fractured Nb:STO surfaces were found scannable by STM in most of the smooth regions seen in SEM image.

\begin{figure}
\includegraphics[width=1\textwidth]{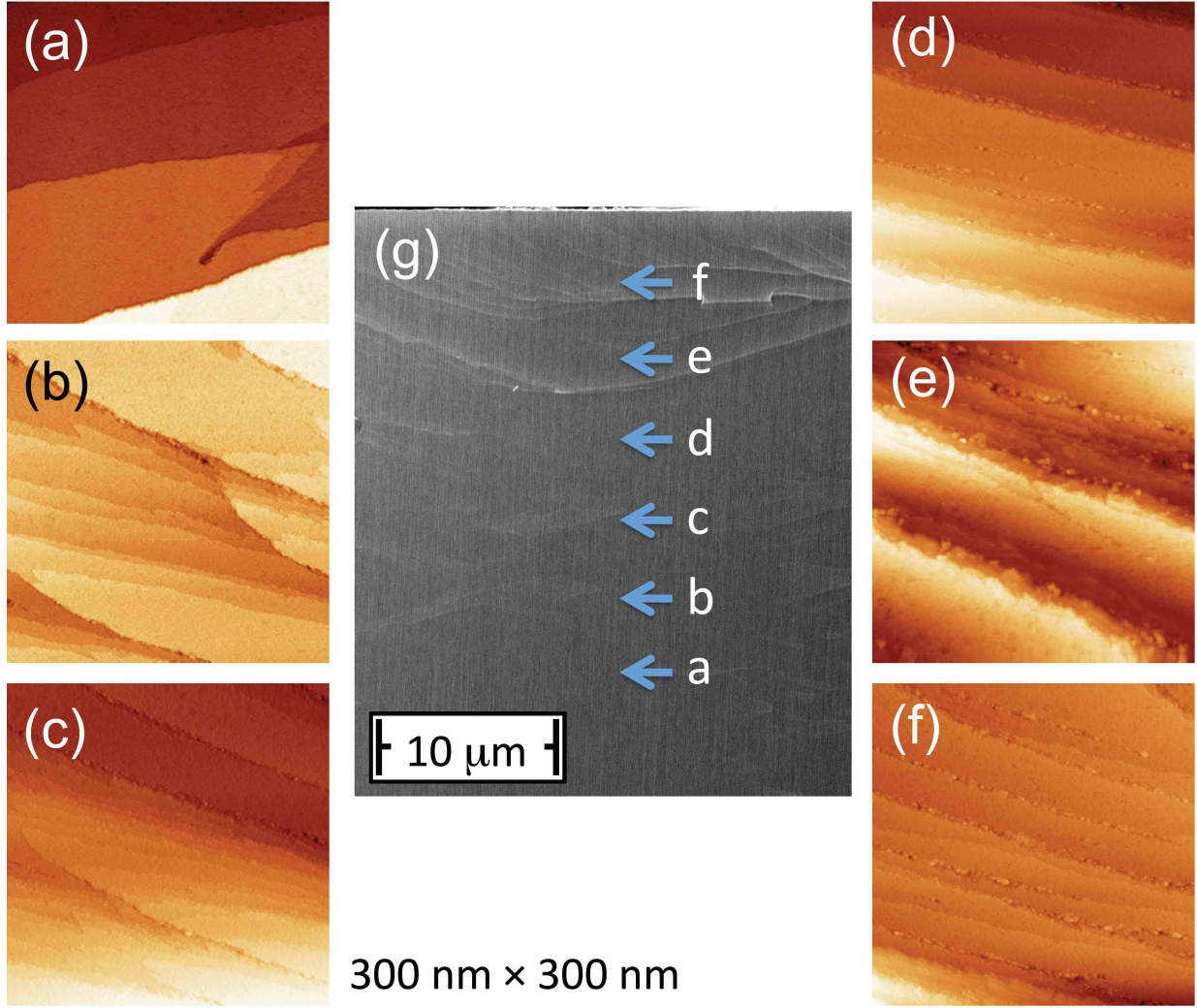}
\caption{\label{fig:line} (a) – (f) Topographies measured by STM at six different positions along [001] direction with $\sim$5 $\mu$m distance apart. Each STM image has dimension of 300 nm $\times$ 300 nm.The measured positions were indicated roughly in the (g) SEM image near sample edge (top of (g)).}
\end{figure}

Figure \ref{fig:line} shows the spatial evolution of the topographies measured by STM along the [001] direction within the red square region indicated in fig. \ref{fig:contour} (b).
Figure \ref{fig:line} (a)-(f) show the topographies with 300 nm $\times$ 300 nm dimensions. Each image was measured $\sim$5 $\mu$m apart along one line toward sample edge ([001] direction), as indicated in fig. \ref{fig:line} (g). Note that since the SEM image was measured \textit{ex-situ}, the arrows in fig. \ref{fig:line} (g) only indicate relative positions of the STM images in (a)-(f). In this case, it can be seen that the topography changes when moving toward the sample edge. In fig. \ref{fig:line} (a), the region furthest from the sample edge, large ($\sim$100 nm) atomically flat terraces were observed. While moving toward the sample edge (fig. \ref{fig:line} (b)-(f)), the terrace width decreases. In addition, wide and narrow terraces appear alternatively along [100] direction. This alternative-terrace behavior was observed in our previous study \cite{guisinger 2009}, which corresponded to alternate surfaces terminations - SrO vs. TiO$_{2}$. This demonstrates that the atomically flat fractured Nb:STO surfaces extended all the way to the sample edge, indicating the study of the interfaces between a thin oxide film and STO is plausible.

\begin{figure}
\includegraphics[width=1\textwidth]{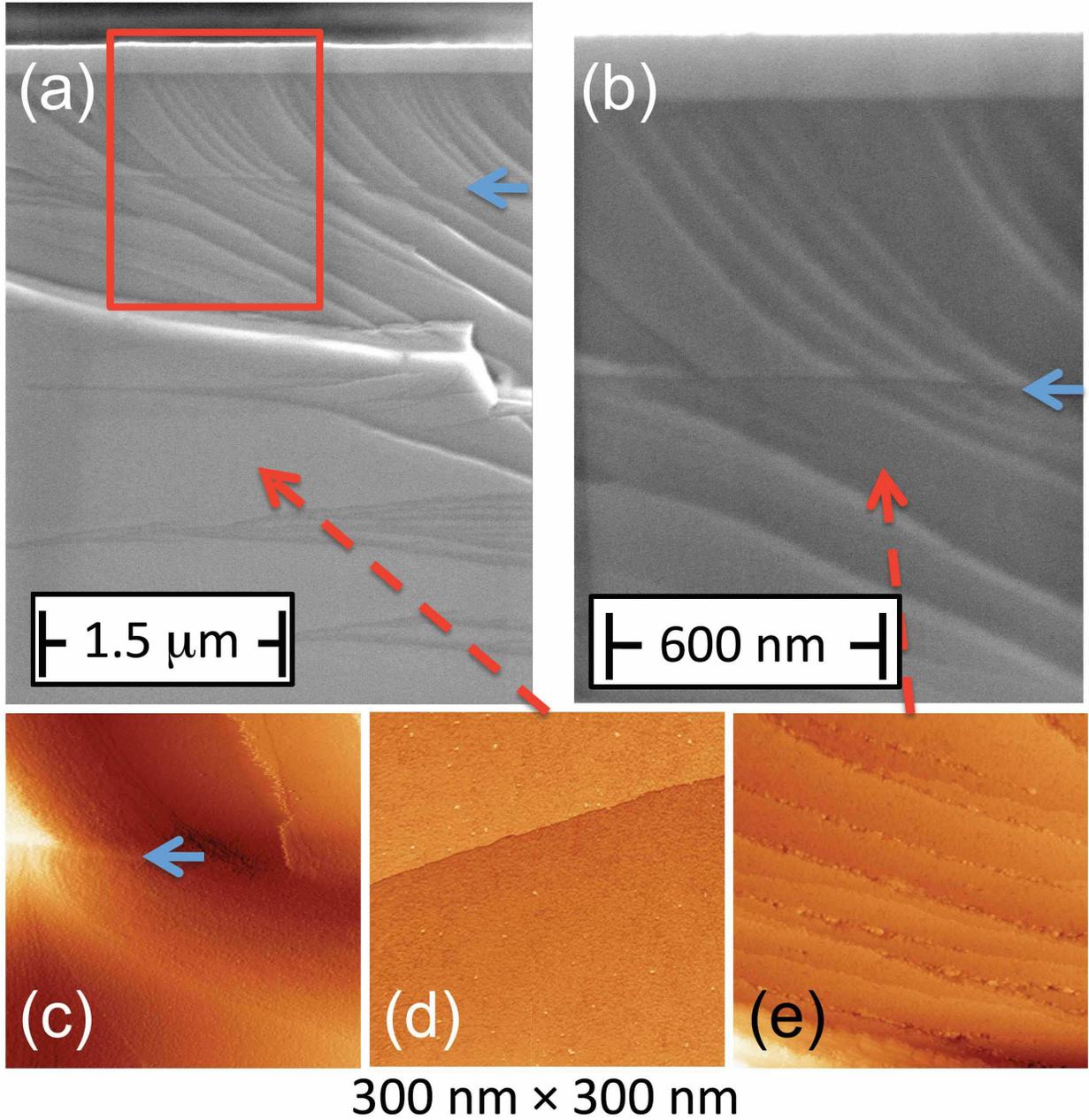}
\caption{\label{fig:dislocation} (a) SEM image showing textures near sample edge; (b) zoom-in SEM image in red rectangular in (a); (c) STM image showing the observation of dislocations (horizontal arrow), which were also seen in (a) and (b); (d) and (e) STM images of corresponding positions indicated with dashed arrows in (a) and (b) respectively. All the STM images, (c)-(e), have the same dimensions of 300 nm $\times$ 300 nm.}
\end{figure}

In addition to the observations mentioned above, dislocations were also observed in both the STM and SEM images. Figure \ref{fig:dislocation} (a) shows a fine scale SEM image with observation of dislocations as indicated by a horizontal arrow.
Figure \ref{fig:dislocation} (b) and (c) show the higher-magnification SEM and STM images, respectively. Dislocations, as indicated with horizontal arrows, were clearly observed. Besides the dislocations, a more dramatic height change was seen in fig. \ref{fig:dislocation} (a) (around the middle of the image). This dramatic height change separates two very different topographies: smooth (below) and textured (above) regions. The STM images scanned in these two regions are shown in fig. \ref{fig:dislocation} (d) and (e), at positions indicated by dashed arrows. Note that, again, the STM arrows do not indicate the exact positions for the STM images, but the relative positions reference to the dramatic height change, which can be sensed by STM. In fig. \ref{fig:dislocation} (d), sub-micron wide terraces were observed; while alternating terminated terraces were found in fig. \ref{fig:dislocation} (e). Unit cell height steps were found near the edge of terraces in fig. \ref{fig:dislocation} (d), indicating the a mixed termination surfaces on the adjacent large terraces, as previously reported by Chien \textit{et al.} \cite{chien 2009}. On the other hand, terraces a few tens of nm wide were found in the region shown to be textured in the SEM images. In this region, two different terminated surfaces were found. In those narrow terrace regions, relatively wider ($\sim$20 nm) and narrower ($\sim$10 nm) terraces were revealed. As reported in our previous study by Guisinger \textit{et al.} \cite{guisinger 2009}, the wide terraces have smoother surfaces (smaller than half unit cell, 0.2 nm) and were assigned to be TiO$_{2}$ terminated surfaces; while the relative narrow terraces have rougher surfaces (about half unit cell), were assigned to be SrO terminated surfaces.

There are many factors that affect the resulting fractured surfaces. For examples, the speed of the fracturing; the sample temperature; the torque during the fracturing etc. The speed of the fracturing in all of our experiments was controlled with the use of a precision motion manipulator mounted in the ultra high vacuum chamber. The samples were sitting on the end of the manipulator and moved with steady speed of about 1 mm/sec. The torque experienced by the samples during the fracturing were minimized by controlling the sample positions, so that the forces applied to the samples were normal to the sample surfaces. We examined the effect of the fracturing temperature at RT and at 50 K. It is found that for RT fractured samples, statistically, narrow and alternating-terminated terraces are the most likely surface; while the samples fractured at 50 K have larger (sub-micron) terrace regions. As discussed above, large and narrow terraces can be found simultaneously on the same fractured surfaces. From this we conclude that the low temperature fracturing must create relatively larger smooth regions with larger terraces than the high temperature fracturing. This is a similar behavior to that observed in layered oxides \cite{pennec 2008}, in which low temperature fracturing created lower defect-density surfaces. In the sense that steps are one kind of defects, our temperature-dependent fracturing on perovskite oxides exhibits similar conclusions. We also note that to we have fractured 14 samples to be examined with XSTM and only two of them were not scannable by STM. This illustrates that even in a complex fracturing process for a non-cleavable material, we can reliably generate atomically flat surfaces to study.

In summary, \textit{in-situ} fracturing was utilized to routinely create atomically flat surfaces from complex oxide perovskite materials. Our survey concluded that the fractured surfaces composed of many different topographic features, from micro-meter scale height change to nano-meter scale flat regions. The scannable region for STM can be found all the way from the center of the fractured surface up to the edge of the sample. This ensures that the study of the pristine perovskite interfaces with thin oxide film overlayers is promising.

Work at Argonne, including the Center for Nanoscale Materials, is supported by the U.S. Department of Energy, Office of Science, Office of Basic Energy Sciences, under Contract No. DE-AC02-06CH11357.

\end{document}